\documentclass[pre,twocolumn,showpacs,preprintnumbers,amsmath,amssymb]{revtex4}

\usepackage{graphicx}
\usepackage{epsfig}
\usepackage{pst-all}
\usepackage{bbm}

\newcommand{\dd}{\mathrm{d}}
\newcommand{\ee}{\mathrm{e}}
\newcommand{\RR}{\mathbbm{R}}

\newcommand{\EE}{\mathbbm{E}}

\renewcommand{\phi}{\varphi}

\providecommand{\norm}[1]{\lVert#1\rVert}

\begin{document}

\title{The mean-field $\boldsymbol{\varphi^4}$-model: entropy, analyticity, and configuration space topology}

\author{Ingo Hahn}
\email{Ingo.Hahn@uni-bayreuth.de}
\affiliation{Physikalisches Institut, Lehrstuhl f\"ur Theoretische Physik I, Universit\"at Bayreuth, 95440 Bayreuth, Germany}

\author{Michael Kastner}
\email{Michael.Kastner@uni-bayreuth.de}
\affiliation{Physikalisches Institut, Lehrstuhl f\"ur Theoretische Physik I, Universit\"at Bayreuth, 95440 Bayreuth, Germany}


\begin{abstract}
A large deviation technique is applied to the mean-field $\varphi^4$-model, providing an exact expression for the configurational entropy $s(v,m)$ as a function of the potential energy $v$ and the magnetization $m$. Although a continuous phase transition occurs at some critical energy $v_c$, the entropy is found to be a real analytic function in both arguments, and it is only the maximization over $m$ which gives rise to a nonanalyticity in $\hat{s}(v)=\sup_m s(v,m)$. This mechanism of nonanalyticity-generation by maximization over one variable of a real analytic function is restricted to systems with long-range interactions and has---for continuous phase transitions---the generic occurrence of classical critical exponents as an immediate consequence. Furthermore, this mechanism can provide an explanation why, contradictory to the so-called topological hypothesis, the phase transition in the mean-field $\varphi^4$-model need not be accompanied by a topology change in the family of constant-energy submanifolds.
\end{abstract}

\pacs{05.70.Fh, 02.40.-k, 64.60.Cn, 75.10.-b}

\maketitle

\section{Introduction}

The theory of phase transitions is an active field of research already since many decades. Despite several remarkable achievements on both the conceptual and the applied side, there still remain open questions galore. As a contribution to the conceptual side, a recent and promising proposal is the {\em topological approach}\/ to phase transitions, connecting the occurrence of a phase transition to certain properties of the potential energy function $v_N$, resorting to topological concepts. Among the remarkable features of this approach we mention:
\begin{enumerate}
\item Topology is in some sense a very reductional level of description. Compared to a geometric description of the same object, lots of information is disregarded in topology. So the topological approach is an attempt to trace back the occurrence of a phase transition to more fundamental quantities than the measures in state space commonly considered.
\item The microscopic Hamiltonian dynamics of a system can be linked via the maximum Lyapunov exponent to topological quantities \cite{CaPeCo:00}. These topological quantities, in turn, are linked to the occurrence of a phase transition by the topological approach. In this way, a  connection is established between phase transitions on the one hand and the underlying microscopic dynamics of the system on the other hand.
\end{enumerate}

The topological approach is based on the hypothesis \cite{CaCaClePe:97} that phase transitions are related to topology changes of a family $\left\{M_v\right\}$ of certain submanifolds $M_v$ of the configuration space of the system. For a system with $N$ degrees of freedom, the $M_v$ consist of all points $\varphi$ of the configuration space $\Gamma$ for which the potential energy $v_N(\varphi)$ per degree of freedom is smaller than or equal to a certain level $v$,
\begin{equation}\label{eq:M_def}
M_v=\left\{ \varphi\in\Gamma\,\big|\,v_N(\varphi)\leqslant v\right\}.
\end{equation}
(Or, in a related version, the topology of submanifolds $\Sigma_v=\left\{ \varphi\in\Gamma\,\big|\,v_N(\varphi)= v\right\}$ is considered.) The hypothesis then conjectures that a topology change within the family $\left\{M_v\right\}$ at $v=v_c$ is a necessary condition for a thermodynamic phase transition to take place at $v_c$ [or at the corresponding critical temperature $T_c=T(v_c)$]. This hypothesis has been corroborated by numerical and exact results for a model showing a discontinuous phase transition \cite{Angelani_etal:03,Angelani_etal:05} as well as for systems with continuous phase transitions \cite{CaPeCo:00,CaCoPe:02,CaPeCo:03,GriMo:04,RiTeiSta:04,Kastner:04,Kastner,KaSchne}. A major achievement in the field is the recent proof of a theorem, stating, loosely speaking, that, for systems described by smooth, finite-range, and confining potentials, a topology change of the submanifolds $M_v$ is a {\em necessary}\/ criterion for a phase transition to take place \cite{FraPe:04,FraPeSpi}.

Recent results on the configuration space topology of the mean-field $\varphi^4$-model have cast some doubt on the general relevance of topology changes for the occurrence of phase transitions. It was first observed in Ref.~\cite{BaroniPhD} that, for this model, the potential energy $v_c$ at which the thermodynamic phase transition takes place does not coincide in general with the potential energy corresponding to any of the topology changes of $\left\{M_v\right\}$. Independently, this result was confirmed in Ref.~\cite{GaSchiSca:04}. The discrepancy between the critical energies from thermodynamics and from topology provoked quite some speculations in the literature \cite{BaroniPhD,AnAnRuZa:04}.

In the present paper, an explanation for this discrepancy is given. The explanation is based on the observation that a topology change in the family $\left\{M_v\right\}$ is {\em one possible}\/ mechanism to entail a thermodynamic singularity. The theorem of Refs.~\cite{FraPe:04,FraPeSpi} then states that for systems with smooth, finite-range, and confining potentials, a topology change is the {\em only}\/ mechanism to generate such a singularity. We individuate a further singularity-generating mechanism, namely a maximization over one of the variables of a real analytic function \footnote{All functions discussed in this article are real, and the term ``analytic'' is used in the sense of real analytic or smooth throughout. By ``nonanalyticity'' we denote a discontinuity of some derivative of arbitrary order of a function.}, which occurs in the mean-field $\varphi^4$-model. The availability of this mechanism accounts for the fact that the phase transition in this model need not be, and in fact is not, related to the topology changes observed. This second singularity-generating mechanism is by no means restricted to the mean-field $\varphi^4$-model, but is genuine to systems with long-range interactions, impossible to take place in short-ranged ones. Not only can this maximization over a real analytic function account for the absence of a relation between thermodynamic and topological quantities, it also leads generically to the so-called mean-field (or classical) critical exponents $\alpha=0$, $\beta=\frac{1}{2}$,... in case of a continuous phase transition.

To the purpose of illustration of this singularity-generating mechanism, we present an analysis of the microcanonical entropy function of the mean-field $\varphi^4$-model, which could stand also as a result of interest on its own. The entropy $s$ is derived from a large deviation principle, a powerful tool from probability theory. As a function of the potential energy $v$ and the magnetization $m$, the entropy is found to be real analytic, and only the maximization over $m$ generates a nonanalyticity in $\hat{s}(v)$.

After fixing some notation in Sec.~\ref{sec:model}, the microcanonical entropy $s$ of the mean-field $\varphi^4$-model is computed in Sec.~\ref{sec:mic_ent}. This result is confronted with the canonical one in Sec.~\ref{sec:equivalence}, where in particular the \mbox{(non-)}equivalence of the statistical ensembles is investigated. In Sec.~\ref{sec:topo}, part of the results on the topology of the $M_v$ from Refs.~\cite{BaroniPhD,GaSchiSca:04,AnAnRuZa:04} is reviewed, and the relation to thermodynamic quantities is discussed. Section \ref{sec:mechanisms} is devoted to two different ways in which a thermodynamic singularity can be generated: by topology changes in the $M_v$, or by maximization over one of the variables of a real analytic function. In the final Sec.~\ref{sec:summary}, we comment on the general applicability of our results to long-range systems, as well as on the altered perspective from which the topological approach to phase transition has to viewed as a consequence.  

\section{Mean-field $\boldsymbol{\varphi^4}$-model}\label{sec:model}

The Ginzburg-Landau or field-theoretic $\varphi^4$-model plays a prominent role in the theory of critical phenomena, mainly due to the fact that it provides a unifying description of various, physically different, lattice and continuum systems near criticality (see \cite{Binney_etal} for an introduction). Here we study a lattice version of this model, where the degree of freedom associated with the $i$-th lattice site is characterized by a scalar variable $\varphi_i\in\RR$. As an analytically tractable caricature of physically relevant long-range forces, we consider mean field-like interactions, where all degrees of freedom interact with each other at equal strength. For such a system consisting of $N$ degrees of freedom, the Hamiltonian function $\mathcal{H}_N:\;\RR^{2N} \to \RR$ is given by
\begin{equation}\label{eq:Hamiltonian}
\mathcal{H}_N(\pi,\varphi)=\frac{1}{2}\sum_{i=1}^N \pi_i^2+N\left[v_N(\varphi)\right],
\end{equation}
where $J$ is a coupling constant, and $\pi=(\pi_1,\dotsc,\pi_N)$ is the vector of momenta conjugate to $\varphi=(\varphi_1,\dotsc,\varphi_N)$. The potential energy is given by
\begin{equation}\label{eq:v_N}
v_N(\varphi)= z_N(\varphi)-\tfrac{J}{2}\left[m_N(\varphi)\right]^2,
\end{equation}
with on-site potential
\begin{equation}\label{eq:z_N}
z_N(\varphi)=\frac{1}{N}\sum\limits_{i=1}^N \left( -\frac{1}{2}\varphi_i^2 + \frac{1}{4}\varphi_i^4 \right)
\end{equation}
and magnetization
\begin{equation}\label{eq:magnetization}
m_N(\varphi)=\frac{1}{N}\sum\limits_{i=1}^N \varphi_i,
\end{equation}
where $v_N$, $z_N$, and $m_N$ are functions from $\RR^N$ onto the reals. Typically, a coupling term $-hm(\varphi)$ to a symmetry breaking magnetic field $h$ is added to the Hamiltonian (\ref{eq:Hamiltonian}). Since we are interested in the zero-field situation, such a term will here be omitted.

Canonical thermodynamic quantities of the mean-field $\varphi^4$-model have been computed in Refs.~\cite{OvOn:90,DauxLeRu:03}. We will make reference to these results when discussing ensemble \mbox{(non-)}equivalence in Sec.~\ref{sec:equivalence}.

\section{Microcanonical ensemble}\label{sec:mic_ent}

In the microcanonical ensemble, the natural starting point for thermodynamics is the entropy. For our purposes, we are interested in the thermodynamic limit of the {\em configurational}\/ entropy, i.\,e., the contribution of the kinetic term in (\ref{eq:Hamiltonian}) to the entropy is disregarded. We will compute and discuss two different---but closely related---entropy functions, the first being a function of the potential energy and the magnetization,
\begin{equation}\label{eq:s_vm}
s(v,m)=\!\lim_{N\to\infty}\frac{1}{N}\ln\!\int_{\RR\!^N}\!\dd\varphi\,\delta\left[v-v_N(\varphi)\right]\delta\left[m-m_N(\varphi)\right],
\end{equation}
where $\delta$ denotes the Dirac distribution. The second entropy function we consider depends only on the potential energy,
\begin{equation}\label{eq:s_v}
\hat{s}(v)=\lim_{N\to\infty}\frac{1}{N}\ln\int_{\RR\!^N}\dd\varphi\,\delta\left[v-v_N(\varphi)\right],
\end{equation}
and can be obtained from $s(v,m)$ by maximization,
\begin{equation}\label{eq:s_maximization}
\hat{s}(v)=\sup_m s(v,m).
\end{equation}
This leaves the calculation of $s(v,m)$ as the main task of this section. Before doing so, we want to say a few words about how a phase transition manifests itself in the microcanonical entropy.

\subsection{Microcanonical entropy and phase transitions}

In the canonical ensemble, a possible (and commonly accepted) definition of a phase transition is the presence of a nonanalyticity in the canonical free energy. In the microcanonical ensemble, conditions on the microcanonical entropy $s$ for the existence of a phase transition are somewhat less established. An obvious way to start is to look for conditions on $s$ which, in the case of ensemble equivalence, correspond to the definition of a phase transition from the canonical free energy density. Such conditions are
\begin{enumerate}
\item the existence of an open subset of the domain of $s$ on which the entropy is not strictly concave, or
\item a nonanalyticity in $s$.
\end{enumerate}
Then it might appear reasonable to adopt the same conditions also for systems for which equivalence of ensembles does not hold. So in order to identify phase transitions in the microcanonical ensemble, we will focus on concavity and analyticity properties of the entropy in the following.

\subsection{Large deviation theory and microcanonical entropy}\label{sec:largedeviation}

Large deviation theory is a branch of probability theory which is concerned with events of very low probability. This section provides an informal introduction as to how large deviation theory can be applied to calculate the microcanonical entropy.

Let $X^N=(X_1,X_2,\dotsc,X_N)$ be a sequence of $N$ independent and identically distributed random variables $X_i\in\RR^n$ with empirical mean
\begin{equation}
\mathfrak{S}_N\left(X^N\right)=\frac{X_1+X_2+\dotsb+X_N}{N}
\end{equation}
and mean $\mu=\EE[X_i]$, where $\EE[\cdot]$ denotes the expectation value. The strong law of large numbers then states that the probability density to find $\mathfrak{S}_N\ne\mu$ converges to zero as $N$ goes to infinity. The theory of large deviations deals with the form of this convergence. Provided existence of the generating function $\EE[\ee^{\langle t,X\rangle}]$, the rate function $I$ is defined as the Legendre-Fenchel transform
\begin{equation}\label{eq:ratefunction}
I(x)=\sup_{t\in\RR^n}\left[\langle t,x\rangle-\ln\EE[\ee^{\langle t,X\rangle}]\right],
\end{equation}
where $\langle\cdot,\cdot\rangle$ denotes the Euclidean scalar product. Then Cram\'er's theorem (Theorem I.4 in \cite{Hollander}) states that the probability $P$ to find the empirical mean $\mathfrak{S}_N\in [x,x+\dd x]$ converges exponentially,
\begin{equation}
P(\mathfrak{S}_N\in [x,x+\dd x])= \ee^{-NI(x)}\,\dd x
\end{equation}
in the limit $N\to\infty$, or, equivalently,
\begin{equation}
-I(x)=\lim_{N\to\infty}\frac{1}{N}\,\ln\,P(\mathfrak{S}_N\in [x,x+\dd x]),
\end{equation}
and $\mathfrak{S}_N$ is said to satisfy a large deviation principle

The connection to statistical physics is made by observing that the microcanonical entropy of a system with $N$ degrees of freedom, described by generalized coordinates $X_i\in\RR^n$, is defined as
\begin{equation}\label{eq:s_N_general}
s_N(x)=\frac{1}{N}\ln \int_{\RR^{Nn}}\dd X^N\, \delta\left[x-\mathfrak{S}_N\left(X^N\right)\right].
\end{equation}
The integral in this equation quantifies the volume in state space $\RR^{Nn}$ occupied by microstates $X^N=(X_1,X_2,\dots,X_N)$ which are compatible with a certain macrostate $x$. If $P(\mathfrak{S}_N\in [x,x+\dd x])$ is the probability to find the system in a state with $\mathfrak{S}_N\in [x,x+\dd x]$, the proportionality of this integral to $P(\mathfrak{S}_N\in [x,x+\dd x])$ follows immediately from the assumption of equal {\em a priori}\/ probability of the microstates, which is at the very basis of equilibrium statistical mechanics. The proportionality constant results in a physically irrelevant summand in the entropy and is omitted, leading to the expression 
\begin{equation}
s(x)=\lim_{N\to\infty}\frac{1}{N}\,\ln\,P(\mathfrak{S}_N\in [x,x+\dd x])
\end{equation}
for the entropy in the limit of large $N$. Provided our problem satisfies a large deviation principle, the microcanonical entropy $s(x)$ for $N\to\infty$ is identical to the negative rate function $-I(x)$ from Eq.~(\ref{eq:ratefunction}). Thus, the high-dimensional integral in (\ref{eq:s_N_general}) is reduced to one single integration [the expectation value $\EE[\ee^{\langle t,X\rangle}]$ in (\ref{eq:ratefunction})] and a maximization,
\begin{equation}\label{eq:s_sup}
s(x)=-\sup_{t\in\RR^n}\left[\langle t,x\rangle-\ln\EE[\ee^{\langle t,X\rangle}]\right].
\end{equation}

$\ln\EE[\ee^{\langle t,X\rangle}$ is strictly convex and infinitely many times differentiable (see proof of Theorem I.4 in \cite{Hollander}). Therefore, there is at most one maximizer $t_x$, and the derivative of the argument of the supremum in (\ref{eq:s_sup}) has to be zero. As a consequence, $t_x$ is determined by the equation 
\begin{equation}\label{legendrefenchelmaximum}
x\EE[\ee^{\langle t_x,X\rangle}] =\EE\left[X\,\ee^{\langle t_x,X\rangle}\right],
\end{equation}
and the maximization can be rewritten as
\begin{equation}\label{entropiex}
s(x)=-\langle t_x,x\rangle+\ln\EE[\ee^{\langle t_x,X\rangle}].
\end{equation}
Note that $-I(x)$, and therefore $s(x)$, are smooth and strictly concave on their domain (Lemma I.14 in \cite{Hollander}).

In \cite{Touchette,BaBoDaRu}, a more extensive discussion of the application of large deviation theory in statistical physics, including a detailed account of the properties of the rate function $I(x)$, is given in a language well accessible to readers with a physics background. A mathematical treatment of large deviation theory can be found in \cite{Hollander,Ellis}.

\subsection{The $\boldsymbol{\phi^4}$-model without interaction}

We are interested in the entropy $s(v,m)$ as a function of the potential energy $v$ and the magnetization $m$ as defined in (\ref{eq:s_vm}). As a first step, applying the concepts of large deviation theory, we compute a related entropy function $\tilde{s}(z,m)$ in this section, where $z$ is the variable associated with the on-site potential $z_N$ defined in (\ref{eq:z_N}). From $\tilde{s}(z,m)$, we then obtain the desired $s(v,m)$ by a simple transformation of variables. In the notation of Sec.~\ref{sec:largedeviation}, we choose $x=(z,m)\in\RR^2$. The adequate empirical mean $\mathfrak{S}_N=\left(z_N(\phi),m_N(\phi)\right)$ is made up from (random) microscopic variables $X_i=\left(\frac{1}{4}\phi_i^4-\frac{1}{2}\phi_i^2,\phi_i\right)$. We further assume uniformly distributed particles with probability density $p(\phi_i=\phi)=\frac{1}{2\phi_c} \;\forall\phi\in[-\phi_c,\phi_c]$ and $0$ elsewhere. Then, the conditional equation (\ref{legendrefenchelmaximum}) for $t_x=(t_z,t_m)$ leads to
\begin{equation}\label{tvtmcondition}
\begin{split}
z&=\tfrac{1}{4}\,\omega_4(t_z,t_m)-\tfrac{1}{2}\,\omega_2(t_z,t_m)\\
m&=\omega_1(t_z,t_m)
\end{split}
\end{equation}
with parameter integrals
\begin{equation}\label{eq:omega_k}
\omega_k(t_z,t_m)=\frac{\int_{-\phi_c}^{+\phi_c} \dd\phi\, \phi^k\,\ee^{t_m \phi + t_z \left(\frac{1}{4}\phi^4 - \frac{1}{2}\phi^2\right)}}{\int_{-\phi_c}^{+\phi_c} \dd\phi\, \ee^{t_m \phi + t_z \left(\frac{1}{4}\phi^4 - \frac{1}{2}\phi^2\right)}} .
\end{equation}
Inserting the solution $t_z(z,m)$, $t_m(z,m)$ of the system of equations (\ref{tvtmcondition}) into (\ref{entropiex}) yields the microcanonical entropy for the noninteracting $\phi^4$-model,
\begin{multline}\label{entropietilde}
\tilde{s}(z,m)=-t_z(z,m)z-t_m(z,m)m\\
+\ln\int_{-\phi_c}^{+\phi_c} \dd\phi\, \ee^{t_m \phi + t_z \left(\frac{1}{4}\phi^4 - \frac{1}{2}\phi^2\right)},
\end{multline}
where we let $\phi_c$ tend to infinity in the following.

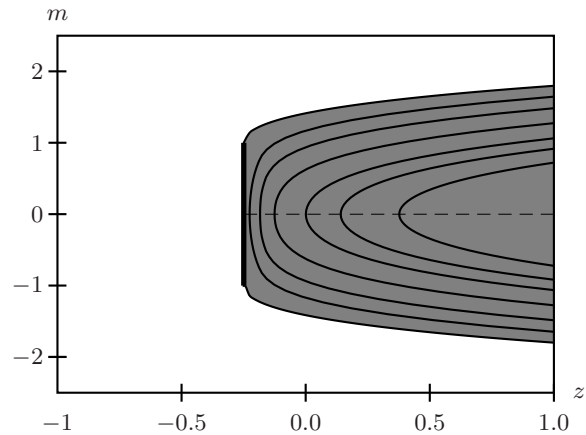
\begin{figure}[ht]
\begin{center}
\psset{xunit=3.3cm,yunit=0.95cm}
  \begin{pspicture}(-1.2,-2.7)(1.1,3)

  \psset{linewidth=0.8pt,linestyle=solid}
  \pscustom{
  \psplot[plotstyle=curve]{1}{-0.25}{x 4 mul 1 add sqrt 1 add sqrt}
  \psline(-0.25,-1)
  \psplot[plotstyle=curve]{-0.25}{1}{x 4 mul 1 add sqrt 1 add sqrt -1 mul}
  \gsave
     \fill[fillstyle=solid,fillcolor=gray]
  \grestore}

  \fileplot[plotstyle=line]{PlotDaten/contoursz1.dat}
  \fileplot[plotstyle=line]{PlotDaten/contoursz2.dat}
  \fileplot[plotstyle=line]{PlotDaten/contoursz3.dat}
  \fileplot[plotstyle=line]{PlotDaten/contoursz4.dat}
  \fileplot[plotstyle=line]{PlotDaten/contoursz5.dat}
  \fileplot[plotstyle=line]{PlotDaten/contoursz6.dat}

  \psline[linewidth=2pt](-0.25,-1)(-0.25,1)

  \psaxes[ticks=x,labels=x,Dx=0.5,Ox=-1,axesstyle=frame](-1,-2.5)(-1,-2.5)(1,2.5)
  \psaxes[ticks=none,labels=y,Oy=-2,Dy=1,axesstyle=none](-1,-2)(-1,-2)(1,2)
  \psaxes[ticks=y,labels=none,Oy=-2,Dy=1,axesstyle=none](-1,-3)(-1,-2)(1,2)

  \psplot[plotstyle=curve,linestyle=dashed,linewidth=0.1pt]{-0.25}{1}{0}

  \rput(1.1,-2.5){$z$}
  \rput(-1,2.8){$m$}

 \end{pspicture}
\end{center}
\caption{\label{figuresmz}
Contour plot of the entropy $\tilde{s}(z,m)$ on its domain (gray) from a numerical evaluation of (\ref{entropietilde}). The entropy is a concave function and minimal on its boundary for maximum $m$. Note that the bold line at $z=-\frac{1}{4}$ is not a contour
.}
\end{figure}

An analytic solution of (\ref{tvtmcondition}), and therefore of the entropy $\tilde{s}$ in (\ref{entropietilde}), is not possible in general. Instead, we will state---without proof---some of the properties of $\tilde{s}$. From the definition (\ref{eq:z_N}) of the on-site potential one obtains immediately $z\in[-\frac{1}{4},\infty)$ for the range of $z$. For any fixed $z\geqslant-\frac{1}{4}$, values in the range $\left[-m_{\text{max}}(z),+m_{\text{max}}(z)\right]$ are accessible for the magnetization $m$. The extremal values $m_{\text{max}}(z)$ are defined by the equation
\begin{equation}
z=\frac{1}{4}m_{\text{max}}^4-\frac{1}{2}m_{\text{max}}^2
\end{equation}
and correspond to states in which all degrees of freedom $\phi_i$ take on the same value. This characterizes the domain of $\tilde{s}$ as illustrated in Fig.~\ref{figuresmz}.

According to the previous section, $\tilde{s}(z,m)$ is strictly concave and analytic on its domain. Taking into account the symmetry of the model [invariance of $z_N(\varphi)$ under substitution $\varphi\to-\varphi$], it follows that $\tilde{s}(z,m)$ is maximal at $m=0$ for any given $z\geqslant-\frac{1}{4}$ (see Fig.~\ref{figuresmz} for an illustration).

\subsection{Microcanonical entropy of the mean-field $\boldsymbol{\phi^4}$-model}\label{sec:entropy}
So far we have considered the mean-field $\phi^4$-model without interparticle interaction. Turning on this interaction changes the potential energy of the system from $z_N$ to $v_N=z_N-\frac{J}{2}m_N^2$. This relation allows to derive the desired entropy function
\begin{equation}\label{entropie}
s(v,m)=\tilde{s}\left(v+\tfrac{J}{2}\,m^2,m\right)
\end{equation}
of the interacting system from the above computed $\tilde{s}(z,m)$ by a simple transformation of variables.

The smooth transformation of variables in (\ref{entropie}) retains the analyticity of $\tilde{s}(z,m)$, so $s(v,m)$ is again an analytic function. In order to see whether the concavity property is conserved as well, we distinguish the two cases of, respectively, negative and positive coupling $J$.

\paragraph*{$J<0$:} Let $\mathfrak{H}_f$ denote the Hessian of some function $f$. Using (\ref{entropie}), we can calculate the determinant
\begin{equation}
\det\mathfrak{H}_s=\det\mathfrak{H}_{\tilde{s}}+J\,\frac{\partial^2\,\tilde{s}(z,m)}{\partial\,z^2}\,\frac{\partial\,\tilde{s}(z,m)}{\partial\,z}.
\end{equation}
The concavity of $\tilde{s}$ implies that $\det\mathfrak{H}_{\tilde{s}}>0$ and $\frac{\partial^2\,\tilde{s}(z,m)}{\partial\,z^2}<0$. If we further take into account that $\frac{\partial\,\tilde{s}(z,m)}{\partial\,z}=-t_z>0$, it follows that $\det\mathfrak{H}_s>0$ for $J<0$, therefore $\mathfrak{H}_s$ is negative definite and $s(v,m)$ is again a strictly concave function. From the symmetry of the model it follows that, for any fixed value of $v$, there is a unique maximum at $m=0$. Then the entropy as a function of $v$ only, $\hat{s}(v)=\sup_{m}s(v,m)=s(v,0)$, is likewise analytic and concave, and no phase transition is found to take place for anti-ferromagnetic coupling $J<0$. In the following we will focus on the more interesting ferromagnetic case. 

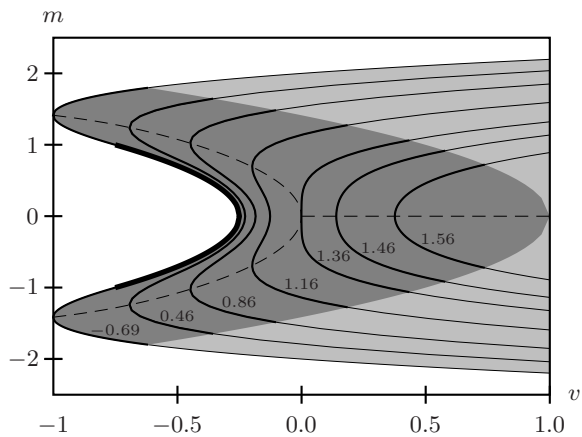
\begin{figure}[ht]
\begin{center}
\psset{xunit=3.3cm,yunit=0.95cm}
  \begin{pspicture}(-1.1,-2.7)(1.1,3)

  \psset{linewidth=0.1pt,linestyle=solid}
  \pscustom{
  \psplot[plotstyle=curve]{1}{-1}{x 1 add sqrt 2 mul 2 add sqrt}
  \psplot[plotstyle=curve]{-1}{-0.75}{x 1 add sqrt -2 mul 2 add sqrt}
  \psplot[plotstyle=curve]{-0.75}{-0.25}{x -2 mul 0.5 sub sqrt}
  \psplot[plotstyle=curve]{-0.2501}{-0.75}{x -2 mul 0.5 sub sqrt -1 mul}
  \psplot[plotstyle=curve]{-0.75}{-1}{x 1 add sqrt -2 mul 2 add sqrt -1 mul}
  \psplot[plotstyle=curve]{-1}{1}{x 1 add sqrt 2 mul 2 add sqrt -1 mul}
  \gsave
     \fill[fillstyle=solid,fillcolor=lightgray]
  \grestore}

  \psset{linewidth=0.8pt,linestyle=solid}
  \pscustom{
  \psplot[plotstyle=curve]{-0.618}{-1}{x 1 add sqrt 2 mul 2 add sqrt}
  \psplot[plotstyle=curve]{-1}{-0.75}{x 1 add sqrt -2 mul 2 add sqrt}
  \psplot[plotstyle=curve]{-0.75}{-0.25}{x -2 mul 0.5 sub sqrt}
  \psplot[plotstyle=curve]{-0.2501}{-0.75}{x -2 mul 0.5 sub sqrt -1 mul}
  \psplot[plotstyle=curve]{-0.75}{-1}{x 1 add sqrt -2 mul 2 add sqrt -1 mul}
  \psplot[plotstyle=curve]{-1}{-0.618}{x 1 add sqrt 2 mul 2 add sqrt -1 mul}
  \gsave
     \psplot[liftpen=1]{-0.618}{1}{-2 x mul 2 add sqrt -1 mul}
     \psplot[liftpen=1]{1}{-0.618}{-2 x mul 2 add sqrt}
     \fill[fillstyle=solid,fillcolor=gray]
  \grestore}

  \fileplot[plotstyle=line,linewidth=0.1pt]{PlotDaten/contoursv1a.dat}
  \fileplot[plotstyle=line,linewidth=0.1pt]{PlotDaten/contoursv2a.dat}
  \fileplot[plotstyle=line,linewidth=0.1pt]{PlotDaten/contoursv3a.dat}
  \fileplot[plotstyle=line,linewidth=0.1pt]{PlotDaten/contoursv4a.dat}
  \fileplot[plotstyle=line,linewidth=0.1pt]{PlotDaten/contoursv5a.dat}
  \fileplot[plotstyle=line,linewidth=0.1pt]{PlotDaten/contoursv6a.dat}

  \fileplot[plotstyle=line,linewidth=0.8pt]{PlotDaten/contoursv1b.dat}
  \fileplot[plotstyle=line,linewidth=0.8pt]{PlotDaten/contoursv2b.dat}
  \fileplot[plotstyle=line,linewidth=0.8pt]{PlotDaten/contoursv3b.dat}
  \fileplot[plotstyle=line,linewidth=0.8pt]{PlotDaten/contoursv4b.dat}
  \fileplot[plotstyle=line,linewidth=0.8pt]{PlotDaten/contoursv5b.dat}
  \fileplot[plotstyle=line,linewidth=0.8pt]{PlotDaten/contoursv6b.dat}

  \psplot[plotstyle=curve,linewidth=2pt]{-0.75}{-0.25}{x -2 mul 0.5 sub sqrt}
  \psplot[plotstyle=curve,linewidth=2pt]{-0.25}{-0.75}{x -2 mul 0.5 sub sqrt -1 mul}
  \psline[linewidth=2pt](-0.25,-0.01)(-0.25,0.01)

  \psplot[plotstyle=curve,linestyle=dashed,linewidth=0.1pt]{0}{1}{0}
  \fileplot[plotstyle=line,linestyle=dashed,linewidth=0.1pt]{PlotDaten/entropiemaximum.dat}

  \psaxes[ticks=x,labels=x,Dx=0.5,Ox=-1,axesstyle=frame](-1,-2.5)(-1,-2.5)(1,2.5)
  \psaxes[ticks=none,labels=y,Oy=-2,Dy=1,axesstyle=none](-1,-2)(-1,-2)(1,2)
  \psaxes[ticks=y,labels=none,Oy=-2,Dy=1,axesstyle=none](-1,-3)(-1,-2)(1,2)

  \rput(1.1,-2.5){$v$}
  \rput(-1,2.8){$m$}

  \rput(-0.75,-1.6){$\scriptstyle-0.69$}
  \rput(-0.5,-1.4){$\scriptstyle0.46$}
  \rput(-0.25,-1.2){$\scriptstyle0.86$}
  \rput(-0.0,-0.95){$\scriptstyle1.16$}
  \rput(0.13,-0.55){$\scriptstyle1.36$}
  \rput(0.31,-0.45){$\scriptstyle1.46$}
  \rput(0.55,-0.32){$\scriptstyle1.56$}

 \end{pspicture}
\end{center}
\caption{\label{figuresmv}
Contour plot of the entropy $s(v,m)$ for coupling constant $J=1$. The gray (light and dark) hatched area is the domain of $s$, dark gray is used for what is obtained from a deformation (variable transform) of the part visible in Fig.~\ref{figuresmz}. The dashed line marks the position of the maximum of $s$ with respect to $m$ for every fixed value of $v$.
}
\end{figure}

\paragraph*{$J>0$:} For arbitrary positive coupling $J$, the domain of $s$ is no more a convex set (see Fig.~\ref{figuresmv}), so $s$ is not a concave function, and we expect a phase transition to take place.

Numerically, we find that for fixed values of $v$ {\em above}\/ a critical value $v_c$, $s(v,m)$ attains its maximum at magnetization $m=0$, whereas {\em below}\/ $v_c$ there is a minimum at $m=0$ and two maxima occur at nonvanishing values of $m$ (Fig.~\ref{figuresmv}). In order to track down the transition point, we expand the entropy $s$ in $m$ around $m=0$, yielding
\begin{multline}\label{entropyexpansion}
s(v,m)=s(v,0)\\-\frac{1}{2}\,\left( J\,\tau_z(v,0) + \frac{1}{\omega_2\left(\tau_z(v,0),0\right)}\right)\,m^2 +{\mathcal O}(m^4),
\end{multline}
where $\tau_z(v,m)=t_z(v+\frac{J}{2}m^2,m)$. Here, ${\mathcal O}$ denotes Landau's order symbol, and the integral
\begin{equation}\label{besselintegral}
\omega_2(\tau_z,0)=\frac{1}{2}\left(1+\frac{I_{-\frac{3}{4}}(-\frac{\tau_z}{8})+I_{\frac{3}{4}}(-\frac{\tau_z}{8})}{I_{-\frac{1}{4}}(-\frac{\tau_z}{8})+I_{\frac{1}{4}}(-\frac{\tau_z}{8})}\right)
\end{equation}
as defined in (\ref{eq:omega_k}) can be expressed in terms of modified Bessel functions of the first kind $I_k$. The transition point is marked by a change of sign of the coefficient of the quadratic term in (\ref{entropyexpansion}), so the critical energy is given via Eq.~(\ref{tvtmcondition}) by $v_c=z(\tau_{z_c},0)$, where $\tau_{z_c}$ is determined implicitly by
\begin{equation}\label{eq:tau_z_c}
J\tau_{z_c} \omega_2\left(\tau_{z_c},0\right)= -1.
\end{equation}
Making use of this equation and the identity
\begin{equation}
\omega_4(t_z,0)=\omega_2(t_z,0)-\frac{1}{t_z},
\end{equation}
the critical energy can be expressed as
\begin{equation}\label{criticalenergy}
v_c(J) =\frac{\frac{1}{J}-1}{4\tau_{z_c}(J)}
\end{equation}
where the dependence of $v_c$ and $\tau_{z_c}$ on the coupling constant $J$ has been indicated. We can immediately read off that $v_c(1)=0$ and, since $\tau_{z_c}<0$, that $v_c>0$ for all $J>1$. Solving explicitly for $\tau_{z_c}$ or $v_c$ appears to be infeasible, but we can derive asymptotic expansions for large and small coupling $J$, respectively.

\paragraph*{Small $J>0$:} The asymptotic expansion of (\ref{besselintegral}) for large $-\tau_{z_c}$ yields 
\begin{equation}
\omega_2(\tau_{z_c},0)=\frac{-4\tau_{z_c}-1}{-4\tau_{z_c}+3}+{\cal{O}}(\tau_{z_c}^{-2}).
\end{equation}
Inserting this result into (\ref{eq:tau_z_c}) and (\ref{criticalenergy}), an expansion of the critical energy in the limit of small coupling $J$ can be obtained,
\begin{equation}\label{eq:vc_smallJ}
v_c(J)=-\tfrac{1}{4}+\tfrac{1}{2}\,J+{\mathcal{O}}(J^2).
\end{equation}

\paragraph*{Large $J>0$:} Analogously to the preceding paragraph, expansion of (\ref{eq:tau_z_c}) for small negative $\tau_{z_c}$ yields an expression for the critical energy in the limit of large $J$,
\begin{equation}\label{eq:vc_largeJ}
v_c(J)=a^2 J^2-\left(2\,a^2-\tfrac{1}{4}\right)J+\left(\tfrac{5\,a^2}{4}-\tfrac{3}{8}+\tfrac{1}{64 \,a^2}\right)+{\cal{O}}\left(\tfrac{1}{J}\right),
\end{equation}
with $a=\Gamma(\frac{3}{4})/\Gamma(\frac{1}{4})\approx 0.338$.

\begin{figure}[ht]
\begin{center}
\psset{xunit=1.8cm,yunit=2cm} 
  \begin{pspicture}(-0.3,-0.6)(4.2,2.2) 
  
  \psaxes[Ox=0,Oy=-0.4,Dx=0.5,Dy=0.4,axesstyle=frame](0,-0.4)(0,-0.4)(4.05,1.81)

  \rput(4.24,-0.42){$J$}  
  \rput(0.,1.95){$v_c$}

  \fileplot[plotstyle=dots,dotstyle=*,dotsize=0.06]{PlotDaten/EcJNum1.dat}
  {\lightgray
  \psplot[plotstyle=curve]{0}{4}{x 2 exp 0.1142366451 mul 0.0954267175 sub 0.0215267096 x mul add}
  \psplot[plotstyle=curve]{0}{0.7}{x 2 div 1 4 div sub}}

  \psset{xunit=5.6cm,yunit=12cm}    
  {\scriptsize \psaxes[Ox=0,Oy=-0.25,Dx=0.1,Dy=0.05,axesstyle=frame,tickstyle=top](0.15,0.1)(0.15,0.1)(0.65,0.255)}
  \psset{origin={-0.15,-0.35}}
  \psplot[plotstyle=curve]{0}{0.3}{x 2 div 1 4 div sub}
  \fileplot[plotstyle=dots,dotstyle=*,dotsize=0.02]{PlotDaten/EcJNum2.dat}
  
 \end{pspicture}
\end{center}
\caption{\label{figurecritical}
Critical energy $v_c$ as a function of the coupling constant $J$ from numerical computation (dots), confronted with the two asymptotic expansions (\ref{eq:vc_smallJ}) and  (\ref{eq:vc_largeJ}).
}
\end{figure}
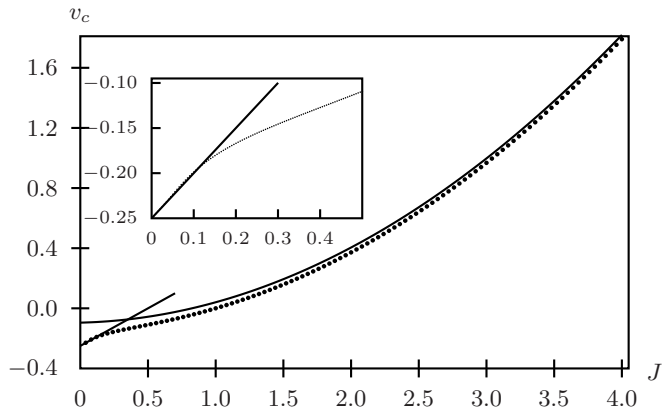

In Fig.~\ref{figurecritical}, the numerically computed behavior of the critical energy $v_c$ as a function of the coupling $J$ is confronted with the asymptotic expansions (\ref{eq:vc_smallJ}) and  (\ref{eq:vc_largeJ}).
Note that, for large enough $J$, the critical energy takes on positive values, as in fact it is divergent in the limit of large $J$. This observation will be of importance for the discussion of the relation between phase transitions and the topology of configuration space submanifolds in Sec.~\ref{sec:topo}.

\subsection{Reduced entropy $\boldsymbol{\hat{s}(v)}$}
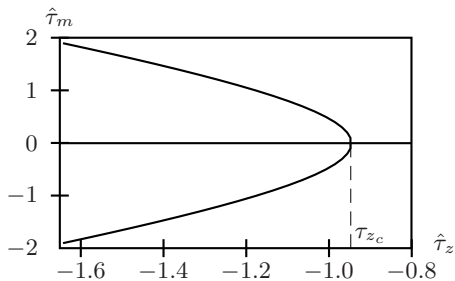
\begin{figure}[ht]
\begin{center}
\psset{xunit=5.5cm,yunit=0.7cm} 
  \begin{pspicture}(-1.8,-2.3)(-0.7,2.7) 
  \psaxes[ticks=y,labels=y,Dy=1,Oy=-2,axesstyle=frame](-1.65,-2)(-1.65,-2)(-0.8,2)
  \psaxes[ticks=none,labels=x,Ox=-1.6,Dx=0.2,axesstyle=none](-1.6,-2)(-1.6,-2)(-0.8,2)
  \psaxes[ticks=x,labels=none,Ox=-2,Dx=0.2,axesstyle=none](-1.8,-2)(-1.8,-2)(-0.8,2)

  \fileplot[plotstyle=line]{PlotDaten/lmlv.dat}
  \psplot[plotstyle=curve]{-1.65}{-0.8}{0}
  \psset{linewidth=0.2pt,linestyle=dashed}
  \qline(-0.947,-2)(-0.947,0)

  \rput(-1.65,2.4){$\hat{\tau}_m$}  
  \rput(-0.72,-1.9){$\hat{\tau}_z$}
  \rput(-0.9,-1.75){$\tau_{z_c}$}

 \end{pspicture}
\end{center}
\caption{\label{fig:maxline}
Numerical solution of the extremalization condition (\ref{conditionmax}) for $J=1$. $\hat{\tau}_m=0$ is a solution for any $\hat{\tau}_z<0$. Below the critical value $\tau_{z_c}$ there are two further solutions.
}
\end{figure}

In Sec.~\ref{sec:entropy}, the entropy $s(v,m)$ was found to be an analytic function. If we release the control parameter $m$, the system will maximize its entropy to $\hat{s}(v)=\sup_{m}s(v,m)$, and we will discuss the analyticity properties of this latter function in the present section. Since $s(v,m)$ is minimal at its boundary, a vanishing first derivative with respect to $m$ is a necessary and sufficient condition for the extremalization. From (\ref{entropietilde}) and (\ref{entropie}) we find the extremizing values of the magnetization $\hat{m}=\omega_1(\hat{\tau}_z,\hat{\tau}_m)$ to be determined by
\begin{equation}\label{conditionmax}
\omega_1(\hat{\tau}_z,\hat{\tau}_m)=-\frac{\hat{\tau}_m}{J\,\hat{\tau}_z}
\end{equation}
The numerical solution of this equation is plotted in Fig.~\ref{fig:maxline}.
As expected from symmetry considerations, Eq.~(\ref{conditionmax}) is trivially solved by $\hat{\tau}_m=0\;\forall\hat{\tau}_z\in\RR^-$, so $\hat{m}=0$ is always an extremum. The Taylor expansion in (\ref{entropyexpansion}) revealed that this solution corresponds to a maximum for $\hat{\tau}_z>\tau_{z_c}$ and to a minimum for $\hat{\tau}_z<\tau_{z_c}$. Numerics ascertain that, for $\hat{\tau}_z>\tau_{z_c}$, the maximum at $\hat{m}=0$ is a global one, and the system is found to be in a paramagnetic phase. For $\hat{\tau}_z<\tau_{z_c}$ two additional solutions $\pm\hat{\tau}_m(\hat{\tau}_z)$ appear. These solutions correspond to two maxima at $\hat{m}=-\frac{\pm\hat{\tau}_m(\hat{\tau}_z)}{J\,\hat{\tau}_z}\ne0$ and reveal the system to be in a ferromagnetic phase. The transition between the two phases is a continuous one (see dashed line in Fig.~\ref{figuresmv}). The critical exponent $\beta$ governing the asymptotic behavior of the magnetization in the vicinity of the transition point can be calculated analytically and is found to be $\beta=\frac{1}{2}$. In the Appendix, we show that the occurrence of classical (=mean-field) values of the critical exponents can be traced back to the analyticity of the entropy function $s(v,m)$, not only in the case of the mean-field $\varphi^4$-model, but generically for mean field-like systems.

In conclusion, the entropy $\hat{s}(v)$ as a function of the potential energy only is given by
\begin{equation}
\hat{s}(v)=\begin{cases}
s(v,0) & \text{for $v > v_c$},\\
s(v,-\frac{\tau_m(\tau_z)}{J\,\tau_z}) & \text{for $v < v_c$},
\end{cases}
\end{equation}
where $\tau_z$ is determined by 
\begin{equation}
v=\tfrac{1}{4}\,\omega_4(\tau_z,\tau_m(\tau_z))-\tfrac{1}{2}\,\omega_2(\tau_z,\tau_m(\tau_z))- \tfrac{[\tau_m(\tau_z)]^2}{2\,J\,\tau_z^2}.
\end{equation}
A nonanalyticity corresponding to a continuous phase transition occurs at $v=v_c$.

\section{Ensemble (non-)equivalence}\label{sec:equivalence}

Results for the mean-field $\varphi^4$-model from computations within the canonical ensemble have been reported in the literature, showing the system to undergo a continuous phase transition \cite{OvOn:90,DauxLeRu:03,AnAnRuZa:04}. Under suitable conditions, the different Gibbs ensembles, like the microcanonical and the canonical one, are known to lead to identical numerical values for the typical system observables of interest in the thermodynamic limit $N\to\infty$. In this case, one speaks of equivalence of ensembles. For a large class of systems with short-ranged interactions, equivalence of ensembles is known to hold \cite{Ruelle,Gallavotti}. A system with mean field-like interactions like the one considered here is clearly not in this class. Nevertheless, comparing the microcanonical result for the critical energy $v_c$ obtained in the previous section with the canonical one reported in \cite{DauxLeRu:03,AnAnRuZa:04}, we observe that the expressions are identical. This is not at all unexpected, as the theorem of G\"artner and Ellis \cite{Ellis} guarantees equivalence of ensembles (on the thermodynamic level we are interested in here, see \cite{ElHaTur} for a precise definition) {\em whenever there is no discontinuous phase transition in the canonical ensemble} \cite{Touchette}.

Despite this ensemble equivalence, the information encoded in the microcanonical results is richer than that of canonical results: The convex region of $s$ (the interior of the dashed parabola-like curve in Fig.~\ref{figuresmv}) is inaccessible within the canonical ensemble, even in the presence of a magnetic field. As a consequence, the analyticity of the microcanonical entropy $s(v,m)$---which will be essential for our discussion of the relation between phase transitions and the topology of configuration space submanifolds---can not be inferred from canonical results.

\section{Topology of configuration space submanifolds}\label{sec:topo}

The topology of the configuration space submanifolds $M_v$ as defined in (\ref{eq:M_def}) has been analyzed for the mean-field $\varphi^4$-model in Refs.~\cite{BaroniPhD,GaSchiSca:04,AnAnRuZa:04} in the framework of Morse theory. To this aim, {\em critical points}\/ \footnote{Not to be confused with thermodynamical critical points.} of the potential energy function (\ref{eq:v_N}) were determined, i.\,e., points $\varphi_c\in\RR^N$ at which the exact differential of $v_N$ vanishes,
\begin{equation}
\dd v_N(\varphi_c)=0.
\end{equation}
For a given coupling constant $J$, all {\em critical values}\/ $v_N(\varphi_c)$ were found to lie within a finite interval,
\begin{equation}
v_N(\varphi_c)\in\left[-\tfrac{1}{4}(J+1)^2,0\right],
\end{equation}
independently of the number $N$ of degrees of freedom. The observation $v_N(\varphi_c)\leqslant0$ is already sufficient for our purposes, as the non-critical neck theorem \cite{Matsumoto} then guarantees that the topology of the $M_v$ remains unchanged for all $v>0$. Recalling the observation from Sec.~\ref{sec:entropy} that the critical potential energy $v_c$ of the phase transition can attain arbitrarily large values, well above the zero upper bound of the critical values of $v_N$, we are led to conclude that the phase transition is found to be uncorrelated with any of the topology changes occurring in the family $\left\{M_v\right\}$ of configuration space submanifolds.

\section{Singularity-generating mechanisms}\label{sec:mechanisms}

The reason why, as observed in the preceding section, the phase transition of the mean-field $\varphi$-model need not be related to the topology changes in the $M_v$ is the existence of a further singularity-generating mechanism. This mechanism is the maximization over the magnetization $m$ in Eq.~(\ref{eq:s_maximization}) of the real analytic entropy function $s(v,m)$, resulting in a nonanalytic $\hat{s}(v)$. In the present section, we will explain why this mechanism is restricted to systems with long-range interactions. Furthermore, it is argued that a topology change in the $M_v$ generates a thermodynamic singularity on a more fundamental level. Although a heuristic argument, we believe that it illustrates the essential difference between the two singularity-generating mechanisms discussed. For the moment, we will restrict ourselves to continuous phase transitions, but modifying the argument as to apply to discontinuous transitions is straightforward. 

In Fig.~\ref{figuresmv}, the typical shape of the entropy $s(v,m)$ of a long-range system in the presence of a symmetry breaking continuous phase transition with some order parameter $m$ is shown: two maxima in $s(v,m)$ with respect to $m$ at potential energies $v$ below the critical energy are opposed to a single maximum at potential energies above the critical energy. As illustrated in Sec.~\ref{sec:mic_ent} for the mean-field $\varphi^4$-model, even an analytic entropy function $s(v,m)$ can entail such a phase transition.

\begin{figure}[ht]
\begin{center}
\psset{xunit=3.3cm,yunit=0.95cm}
  \begin{pspicture}(-1.2,-2.7)(1.2,3)

  \psset{linewidth=0.8pt,linestyle=solid}
  \pscustom{
  \psplot[plotstyle=curve]{1}{-1}{x 1 add sqrt 2 mul 2 add sqrt}
  \psline[linewidth=1pt](-1,-1.41)
  \psplot[plotstyle=curve]{-1}{1}{x 1 add sqrt 2 mul 2 add sqrt -1 mul}
  \gsave
     \fill[fillstyle=solid,fillcolor=lightgray]
  \grestore}

  \fileplot[plotstyle=line,linewidth=0.8pt]{PlotDaten/contoursv1a.dat}
  \fileplot[plotstyle=line,linewidth=0.8pt]{PlotDaten/contoursv2a.dat}
  \fileplot[plotstyle=line,linewidth=0.8pt]{PlotDaten/contoursv3a.dat}
  \fileplot[plotstyle=line,linewidth=0.8pt]{PlotDaten/contoursv4a.dat}
  \fileplot[plotstyle=line,linewidth=0.8pt]{PlotDaten/contoursv5a.dat}
  \fileplot[plotstyle=line,linewidth=0.8pt]{PlotDaten/contoursv6a.dat}

  \psplot[plotstyle=curve,linewidth=2pt]{-0.75}{-0.25}{x -2 mul 0.5 sub sqrt}
  \psplot[plotstyle=curve,linewidth=2pt]{-0.2501}{-0.75}{x -2 mul 0.5 sub sqrt -1 mul}

  \psplot[plotstyle=curve,linestyle=dashed,linewidth=0.1pt]{0}{1}{0}
  \pscustom{
  \gsave
     \fileplot[liftpen=1]{PlotDaten/entropiemaximum.dat}
     \fill[fillstyle=solid,fillcolor=lightgray]
  \grestore}
  \psline[linewidth=0.8pt](-0.69,-1.25)(-0.69,1.25)
  \psline[linewidth=0.8pt](-0.446, -1.03)(-0.446, 1.03)
  \psline[linewidth=0.8pt](-0.198, -0.7166)(-0.198, 0.7166)
  \psline[linewidth=0.8pt](0,-0.15)(0,0.15)

  \psline[linewidth=0.8pt](-1,-1.41)(-1,1.41)
  \fileplot[plotstyle=line,linestyle=dashed,linewidth=0.1pt]{PlotDaten/entropiemaximum.dat}

  \psaxes[ticks=none,labels=x,Dx=0.5,Ox=-1,axesstyle=none](-1,-2.5)(-1,-2.5)(1,2.5)
  \psaxes[ticks=x,labels=none,Dx=0.5,Ox=-1,axesstyle=none](-1.5,-2.5)(-1.5,-2.5)(1,2.5)
  \psaxes[ticks=none,labels=y,Oy=-3,Dy=1,axesstyle=none](-1.1,-3)(-1,-3)(1,2)
  \psaxes[ticks=y,labels=none,Oy=-3,Dy=1,axesstyle=none](-1.1,-3)(-1,-3)(1,2)
  \psaxes[ticks=none,labels=none,axesstyle=frame](-1.1,-2.5)(-1.1,-2.5)(1,2.5)

  \rput(1.1,-2.5){$v$}
  \rput(-1.1,2.8){$m$}

 \end{pspicture}
\end{center}
\caption{\label{figurehull}
Sketch of a typical entropy function $s(v,m)$ of a system with short-range interactions showing a continuous phase transition. The shape is qualitatively that of the concave hull of the entropy of a long-range system (like the one depicted in Fig.~\ref{figuresmv}). The dashed line marks the spontaneous magnetization.
}
\end{figure}
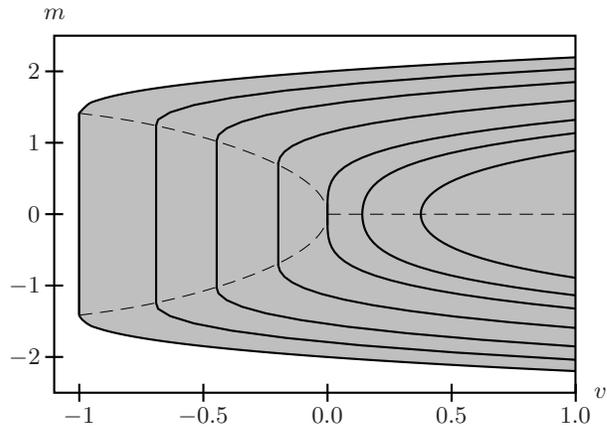

This is different for a system with short-range interactions \footnote{More precisely, the technical conditions on the interaction potential are {\em stability}\/ and {\em temperedness}, see \cite{Gallavotti} for a definition.}. In this case, the entropy as a function of some macroscopic variables $v$, $m$, $x$,... is known to be a concave function \cite{Gallavotti}, and the typical shape of $s(v,m)$ in the presence of a phase transition is obtained by taking the concave hull of a long-range system's entropy (see Fig.~\ref{figurehull} for an illustration). Due to its concavity, the entropy $s$ is nonanalytic not only as a function of $v$ only, but the same holds true for $s(v,m)$ or for $s(v,x)$ for some typical macroscopic quantity $x$.

We will try to trace back the difference between the long-range and the short-range case to the integrals governing the entropy functions. We use a co-area formula from \cite{Federer} to rewrite Eq.~(\ref{eq:s_v}) as
\begin{equation}\label{eq:s_v_coarea}
\hat{s}(v)=\lim_{N\to\infty}\frac{1}{N}\int_{\partial M_v} \frac{\dd \varphi}{\norm{\nabla v_N}}
\end{equation}
with norm
\begin{equation}
\norm{\nabla v_N}=\sqrt{\sum_{i=1}^N\left(\frac{\partial v_N(\varphi)}{\partial \varphi_i}\right)^2}.
\end{equation}
$\partial M_v=\Sigma_v$ denotes the boundary of $M_v$. Similarly, the entropy as a function of $v$ and $x$ [with $x=m$ yielding Eq.~(\ref{eq:s_vm}) as a special case] can be expressed as
\begin{equation}\label{eq:s_vm_coarea}
s(v,x)=\lim_{N\to\infty}\frac{1}{N}\int_{\partial M_v} \frac{\dd \varphi}{\norm{\nabla v_N}}\,\delta\left[x-x_N(\varphi)\right],
\end{equation}
with some smooth function $x_N:\RR^N\to\RR$ corresponding to the macroscopic variable $x$. Then the above discussed analyticity properties of the entropy functions have the following implications for the integrals in (\ref{eq:s_v_coarea}) and (\ref{eq:s_vm_coarea}):

For the case of short-range interactions, we observed above that, in the presence of a phase transition, $\hat{s}(v)$ as well as $s(v,x)$ are necessarily nonanalytic, regardless of the particular form of $x_N$. So if its not the form of the integrand which is crucial, we are led to conjecture that it is the domain of integration $\partial M_v$, and in particular some topology change in the $M_v$, which accounts for the nonanalyticities in $\hat{s}(v)$ and $s(v,x)$. This argument is consistent with the theorem in Refs.~\cite{FraPe:04,FraPeSpi}, proving the necessity of a topology change for a phase transition to take place for a certain class of short-range systems.

For systems with long-range interactions, we found that a nonanalytic $\hat{s}(v)$ can emanate from an analytic $s(v,m)$. In this case, as opposed to our reasoning for short-ranged systems, the emergence of a nonanalyticity in the thermodynamic limit $N\to\infty$ appears to depend on the shape of the integrand, not on the domain of integration $\partial M_v$. This explains on a heuristic level why, for systems with long-range interactions, a phase transition is not necessarily connected to a topology change in the family $\{M_v\}$ of configuration space submanifolds.

The major difference between the two singularity-generating mechanisms is that a topology change in $\{M_v\}$ may lead to a nonanalyticity, regardless of the particular choice of arguments of the entropy function, which is obviously not true for a singularity stemming from a maximization over one of the variables of an analytic entropy function. It is in this sense that by a topology change a nonanalyticity is generated on a more fundamental level.

\section{Summary and conclusions}\label{sec:summary}

Applying a large deviation technique, we have obtained an exact expression for the microcanonical (configurational) entropy $s$ of the mean-field $\varphi^4$-model as a function of the potential energy $v$ and the magnetization $m$. Although the system undergoes a continuous phase transition at some critical energy $v_c$, the entropy $s(v,m)$ is found to be an analytic function in both variables. Only the reduction $\hat{s}(v)=\sup_m s(v,m)$ to an entropy function of one variable gives rise to a nonanalyticity, corresponding to the phase transition of the model.

As expected, the phase transition is found to be governed by classical (=mean-field) critical exponents $\alpha=0, \beta=\frac{1}{2},\dotsc$ The occurrence of these values can be traced back to the analyticity of the entropy $s(v,m)$, not only in the case of the mean-field $\varphi^4$-model, but also generically for systems with mean field-like interactions.

A topology change within the family $\{M_v\}$ of configuration space submanifolds $M_v$ [as defined in (\ref{eq:M_def})] is known to be a mechanism which can give rise to a nonanalyticity in some thermodynamic potential, thus entailing a phase transition. The observed analyticity properties of the entropy functions allow us to individuate the maximization over one variable of an analytic function as a further such mechanism. For the mean-field $\varphi^4$-model, the topology changes were found to be unrelated to the phase transition. This, previously unexpected, observation can be explained by the presence of this second singularity-generating mechanism. Concavity of the entropy of short-range systems restricts the mechanism of singularity generation by maximization to systems with long-range interactions.

We want to conclude with a remark on the implications the results of this article have for what we called the topological approach to phase transitions in the Introduction. The identification of a topology change in the $M_v$ as one out of two, or several, mechanisms to produce a thermodynamic singularity alters profoundly the perspective from which the topological approach has to be viewed. Clearly, in its generality, the above mentioned topological hypothesis \cite{CaCaClePe:97} has to be discarded: in general, a topology change in the $M_v$ at some potential energy $v=v_c$ is not necessary for a phase transition to take place at $v_c$. But this does not diminish the interest in the topological approach. Instead, the classification of phase transitions according to the mechanism by which the thermodynamic singularity is generated opens up as a perspective, with topology changes in configuration space being one---for short-range systems possibly the most prominent one---among these mechanisms. In addition to the two singularity-generating mechanism discussed in this article, there are indications for at least one further such mechanism: also in short-range systems with non-confining potentials, a topology change does not appear to be necessary for the existence of a phase transition \cite{AnRuZa}, while at the same time the maximization mechanism is excluded due to the short-rangedness. One might conjecture that it is the fact that, as a consequence of the potentials being non-confining, the domains of integration in Eqs.~(\ref{eq:s_v_coarea}) and (\ref{eq:s_vm_coarea}) are non-compact manifolds, which might render a third singularity-generating mechanism possible, but this point requires further investigation.

\begin{acknowledgments}
M.\ K.\ acknowledges financial support by the Deutsche Forschungsgemeinschaft (grant KA2272/2).
\end{acknowledgments}

\appendix*

\section{Critical exponents for mean field-like systems}\label{app:expo}

In this appendix, we show that generically (in a sense which will become clear in the following) classical (=mean-field) values follow from the analyticity of the entropy $s$. The mean-field $\varphi^4$-model can then be proved explicitly to be such a generic case by showing that certain coefficients of a Taylor expansion of $s$ do not vanish.

We begin with the remark that the arguments of Sec.~\ref{sec:mic_ent}, guaranteeing the analyticity of the entropy $s(v,m)$, apply not only to the mean-field $\varphi^4$-model, but to general mean field-like systems for which it is possible to express the potential energy by means of macroscopic variables of the kind $\sum_{i=1}^N g_k(\varphi_i)$ with some functions $g_k$ \footnote{For the mean-field $\varphi^4$-model we have $g_1(\varphi_i)=\varphi_i$ and $g_2(\varphi_i)=\frac{1}{4}\varphi_i^4-\frac{1}{2}\varphi_i^2$, yielding the macroscopic variables $m_N$ and $z_N$, respectively.}. In fact, analyticity of the entropy holds even for systems with more realistic long-range forces, whenever the potential energy can be expressed by macroscopic variables plus some remainder, vanishing in the thermodynamic limit (see \cite{BaBoDaRu} for an example).

In order to simplify the presentation, the following derivation is formulated for systems which are symmetric with respect to a (scalar) order parameter $m$, i.\,e., $s(v,m)=s(v,-m)$, but a more general setting is possible. Assuming analyticity, $s(v,m)$ can be expanded into a Taylor series around $m=0$,
\begin{equation}\label{eq:generalexpansion}
s(v,m)=s(v,0)+f_2(v)\,m^2 + f_4(v)\,m^4+ {\mathcal O}(m^6),
\end{equation}
with $f_2(v)=\tfrac{\partial^2 s}{\partial m^2}\big|_{m=0}$ and $f_4(v)=\tfrac{\partial^4 s}{\partial m^4}\big|_{m=0}$.
Odd orders in $m$ vanish due to the system's symmetry. We will understand by the above mentioned {\em generic case}\/ that no coefficients in the expansion (with respect to both, $m$ and $v$) vanish accidentally, so we assume nonzero coefficients throughout.

The extremizing values of the magnetization are determined by $\frac{\partial s}{\partial m}=0$. Applying this to (\ref{eq:generalexpansion}) leads to the trivial solution $\hat{m}=0$ and, for small $\hat{m}\ne 0$, to
\begin{equation}\label{eq:equlibriummagnetization}
\hat{m}(v)\sim\pm\sqrt{-\tfrac{f_2(v)}{f_4(v)}}
\end{equation}
whenever the radical is positive. For the concept of critical exponents to be meaningful, we assume a continuous phase transition to take place. This implies $f_2(v)$ to be zero at some critical point $v=v_c$, and the expansions of $f_2$ and $f_4$ around $v_c$ are of the form 
\begin{eqnarray}
f_2(v)&=&f'_2(v_c)\,(v-v_v)+{\mathcal O}\left((v-v_c)^2\right)\label{eq:f2cond}\\
f_4(v)&=&f_4(v_c)+{\mathcal O}\left(v-v_c\right)\label{eq:f4cond}
\end{eqnarray}
Inserting these expressions into (\ref{eq:equlibriummagnetization}), we obtain
\begin{equation}
\hat{m}(v)\sim\pm\sqrt{-\tfrac{f'(v_c)}{f_4(v_c)}}\sqrt{v-v_c}\propto\left(v-v_c\right)^{1/2}
\end{equation}
in leading order for the equilibrium magnetization in the vicinity of the $v_c$. Similarly, the asymptotic behavior of other thermodynamic quantities can be determined from a generic analytic entropy function (see \cite{KastnerDiss} for derivations). To obtain critical exponents, the proportionalities to $v-v_c$ have to be translated into proportionalities to the reduced temperature $t=\frac{T-T_c}{T_c}$ by means of the asymptotic relation $v-v_c\sim t^{1-\alpha}$, where $\alpha$ is the critical exponent of the specific heat. In this way, we finally obtain a critical exponent $\beta=\frac{1}{2}$ for the equilibrium magnetization of a system with analytic entropy $s$ undergoing a continuous phase transition.

An elaborate but straightforward calculation confirms that, for the mean field $\phi^4$-model, the above conditions (\ref{eq:f2cond}) and (\ref{eq:f4cond}) on $f_2$ and $f_4$ hold with $f'_2,f_4\neq0$, rendering this model a representative of the class of generic systems with classical critical exponents.

\bibliographystyle{h-physrev}
\bibliography{phi4mf}

\end{document}